\documentclass{sig-alternate-per}

\usepackage[utf8]{inputenc}
\usepackage{caption}
\usepackage{amsmath,amssymb,amsfonts}
\usepackage{graphicx}
\usepackage{textcomp}
\usepackage{xcolor}
\usepackage{listings}
\usepackage{graphicx, float}
\usepackage{multicol}
\usepackage{multirow}
\usepackage{etoolbox}
\usepackage{caption}
\usepackage[ruled,vlined]{algorithm2e}
\usepackage{algpseudocode}
\usepackage[hyphens]{url}
\usepackage{balance}
\usepackage{subfig}
\usepackage{adjustbox}

\usepackage[square,numbers]{natbib}

\makeatletter
\renewcommand*{\@biblabel}[1]{\hfill[#1]}
\makeatother

\newif\ifupccolor

\upccolortrue
\begin{document}
% To "force" respecting margins
\sloppy

\title{Unveiling the potential of Graph Neural Networks for robust Intrusion Detection}

%\author{David Pujol-Perich\footnotemark[1]~, Jos\'e Su\'arez-Varela\footnotemark[1]~, Albert Cabellos-Aparicio\footnotemark[1]~, Pere Barlet-Ros\footnotemark[1]\\
%\affaddr{\footnotemark[1]~~Barcelona Neural Networking Center, Universitat Polit\`ecnica de Catalunya, Spain}\\
%\footnotemark[2]Network Technology Lab., Huawei Technologies Co., Ltd.
%\url{contactus@bnn.upc.edu}
%}

\author{David Pujol-Perich, Jos\'e Su\'arez-Varela, Albert Cabellos-Aparicio, Pere Barlet-Ros\\
\affaddr{Barcelona Neural Networking Center, Universitat Polit\`ecnica de Catalunya, Spain}\\
\url{contactus@bnn.upc.edu}
}

%\conferenceinfo{Workshop on AI in Networks and Distributed Systems (WAIN) 2021}{\\Milan,Italy}

\maketitle

\begin{abstract}

The last few years have seen an increasing wave of attacks with serious economic and privacy damages, which evinces the need for accurate Network Intrusion Detection Systems (NIDS). Recent works propose the use of Machine Learning (ML) techniques for building such systems (e.g., decision trees, neural networks). However, existing ML-based NIDS are barely robust to common adversarial attacks, which limits their applicability to real networks. A fundamental problem of these solutions is that they treat and classify flows independently. In contrast, in this paper we argue the importance of focusing on the structural patterns of attacks, by capturing not only the individual flow features, but also the relations between different flows (e.g., the source/destination hosts they share). To this end, we use a graph representation that keeps flow records and their relationships, and propose a novel Graph Neural Network (GNN) model tailored to process and learn from such graph-structured information. In our evaluation, we first show that the proposed GNN model achieves state-of-the-art results in the well-known CIC-IDS2017 dataset. Moreover, we assess the robustness of our solution under two common adversarial attacks, that intentionally modify the packet size and inter-arrival times to avoid detection. The results show that our model is able to maintain the same level of accuracy as in previous experiments, while state-of-the-art ML techniques degrade up to 50\% their accuracy (F1-score) under these attacks. This unprecedented level of robustness is mainly induced by the capability of our GNN model to learn flow patterns of attacks structured as graphs.
\end{abstract}


\section{Introduction}

Recent years have witnessed a great surge of malicious activities on the Internet, leading to major service disruptions and severe economic and privacy implications. For example, according to \cite{cost-1} the average cost of a data breach in 2020 was \$3.86 million, while cyber attacks had an estimated cost to the U.S. economy between \$57 billion and \$109 billion only during 2016. These figures urge the need for the development of effective Network Intrusion Detection Systems (NIDS) that can detect \mbox{-- and} thus \mbox{prevent --} future attacks.

In this context, a recent body of literature proposes the use of Machine Learning (ML) techniques as accurate methods to build NIDS~\cite{khraisat2019survey, resende2018survey}. Indeed, existing solutions often show an accuracy above 98\% when evaluated in popular IDS datasets. However, despite their good performance, these ML techniques have been barely considered for commercial NIDS~\cite{sommer2010outside}. We argue that the main reason behind this lack of adoption is their insufficient robustness against traffic changes, adversarial attacks~\cite{corona2013adversarial}, and generalization over traffic of other networks, which are crucial factors to achieve practical ML-based solutions applicable real networks in production.

A main limitation of existing solutions is that most of them treat and classify flows independently, by capturing meaningful flow-level features that correlate with different attacks. This assumption, however, does not properly adapt to numerous real-world attacks that rely on complex multi-flow strategies (e.g., DDoS, port scans). In this paper, we argue that, to effectively detect this type of attacks, it is essential to capture not only the individual features of flows, but also their relationships within the network. Thus, we propose a graph representation we call \textit{host-connection graphs}, which structures flow relationships in a proper way to then capture meaningful information about the structural flow patterns of attacks (e.g., DDoS, port/network scans, brute force attacks). This is mainly supported by the fact that many common attacks can be unambiguously characterized by structural flow patterns that are fixed by the nature of the attack itself.

In this context, we propose a novel Graph Neural Network (GNN) model that uses a non-standard message-passing architecture specially designed to process and learn from host-connection graphs\footnote{The implementation is open source and publicly available at https://github.com/BNN-UPC/GNN-NIDS using the IGNNITION framework \cite{pujol-perich2021ignnition}.}. GNNs~\cite{scarselli2008graph} are a novel neural network family that is specially suitable for processing information inherently represented as graphs (e.g., chemistry, computer networks, physics). As a result, our model shows good capabilities over the graph-structured information within host-connection graphs.

In the evaluation, we reveal the potential of the proposed GNN model to achieve robust NIDS solutions. First, we evaluate our model in the well-known CIC-IDS2017 dataset~\cite{sharafaldin2018toward}. Our results show that the proposed model is able to accurately detect a wide variety of up-to-date attacks, achieving similar accuracy to state-of-the-art ML techniques widely used for NIDS (0.99 of weighted F1-score). Then, we test the robustness of our solution under different common adversarial attacks~\cite{corona2013adversarial}, which are focused on modifying specific flow features, such as the packet size and the inter-arrival times. The results show that our GNN-based NIDS is completely robust to this type of detection prevention methods commonly used by attackers. In contrast, the state-of-the-art ML benchmarks evaluated significantly decrease their performance, observing degradations of accuracy (F1-score) up to 50\% in our experiments. These results suggest that the proposed GNN model is able to capture meaningful patterns from flow relationships, that are more robust to the adversarial attacks analyzed in this paper.

\section{Why graph-based NIDS?} \label{sec:graph_learning}
\begin{figure}
    \centering
    \includegraphics[width=\linewidth]{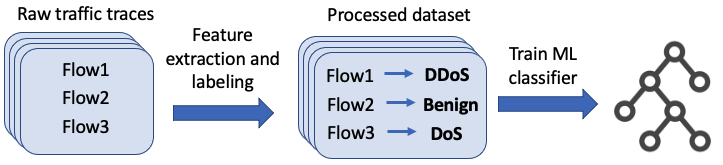}
    \caption{Scheme of traditional ML-based NIDS with flow-based operation.}
    \label{fig:traditional_ml}
    \vspace{-0.3cm}
\end{figure}

Traditionally, ML-based NIDS leverage supervised-learning algorithms, such as Decision Trees, Random Forest, or Support Vector Machines (SVM) to classify the traffic. To train such systems (see Figure~\ref{fig:traditional_ml}), first individual flow records are built from traffic captures, including some features that can be relevant to then classify flows (e.g., packet lengths, inter-arrival times, duration). Afterward, each flow record is labeled according to the attack it represents. Then, a ML model is trained to classify flows individually, based on the information contained in their records.

While these models often achieve good accuracy when trained and evaluated with traffic of the same network, they are specially vulnerable to adversarial attacks, which often vary flow features along time to avoid detection. This limitation becomes particularly evident from the optic of multi-flow attacks, where it is typically needed to analyze and relate a set of flows before detecting the malicious action (e.g., port scans, network scans, DDoS, brute force attacks).

\begin{figure}
    \centering
    \includegraphics[width=\linewidth]{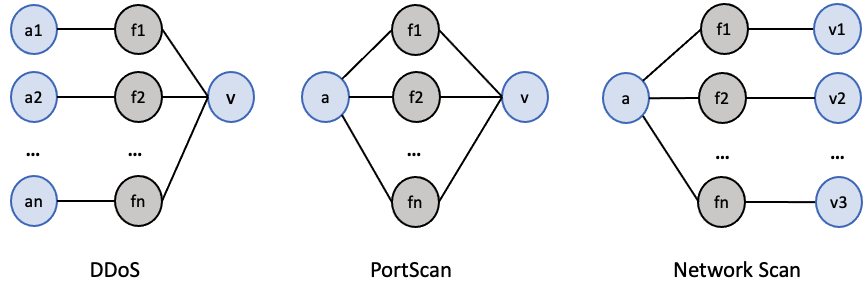}
    \caption{Graph-based representation of well-known attacks. \textit{a$_{x}$} nodes refer to the attackers, \textit{v$_{x}$} nodes represent the victims, and \textit{f$_{x}$} nodes represent different flows.}
    \vspace{-0.2cm}
    \label{fig:graph_patterns}
\end{figure}

Instead, we argue the importance of capturing and modeling the inter-dependencies between different flows traversing the network, which can be naturally represented in the form of graphs. As an example, Figure~\ref{fig:graph_patterns} shows graph representations of common multi-flow attacks. As we can observe, these attacks present inherent flow patterns that make them easily identifiable. For instance, DDoS attacks are distributed by definition, which means that we can expect a massive number of connections \textit{f$_{x}$} from different hosts \textit{a$_{x}$} to the same target \textit{v}. Another classic example are port scans, which involve numerous connections \textit{f$_{x}$} from the same host \textit{a} to different ports of a same destination host \textit{v}. Or network scans, which typically involve multiple connections \textit{f$_{x}$} to hosts of the same network \textit{v$_{x}$} from a single source \textit{a}. In these cases, inspecting flows individually, as most traditional ML-based NIDS (Figure~\ref{fig:traditional_ml}), reasonably hinders the possibility to discriminate such attacks from benign traffic. In practice, some traditional ML-based NIDS have shown high accuracy levels on these attacks, however this can be arguably explained by a high degree of over-fitting on the training and validation datasets, as ML models can eventually learn specific flow-level features (e.g., average packet size) that are highly correlated to some attacks. Nevertheless, this makes them strongly vulnerable to simple variations on malicious flows (e.g., packet lengths, inter-arrival times, ports), which is a common practice among attackers.

In light of the above, we claim that learning the underlying structural flow patterns of attacks is essential to achieve a deeper knowledge and characterization of them, specially for attacks involving multiple flows. More importantly, representing flows and their relations as graphs -- as those of Figure~\ref{fig:graph_patterns} -- enables to capture more robust patterns against potential adversarial attacks, which typically keep the same flow structure, as it is fixed by the nature of the attack itself.

\section{Background on GNN}
\label{sec:background}
Graph Neural Networks (GNN)~\cite{scarselli2008graph} are a recent neural network family specifically designed to learn and generalize over graph-structured data, by capturing and modeling the inherent patterns in graphs. This has resulted in an unprecedented predictive power in many applications where data is structured as graphs~\cite{wu2020comprehensive}. This section describes the basic  architecture of \textit{Message-Passing Neural Network (MPNN)}~\cite{gilmer2017neural}, which represents a general framework covering most of the existing state-of-the-art GNN models~\cite{wu2020comprehensive}.

MPNN operates over a graph \textit{G=(V,E)}, where every node $v \in V$ is characterized with an initial set of features $X_v$, used to encode its initial hidden-state $h_v^0$ (which is represented as a n-element vector). The MPNN then proceeds with the message-passing phase, which is repeated a given number of iterations~$T$. In each message-passing iteration $t$, every node $v$ receives a \textit{message} $m_{v,j}$ from each of its neighbors $j \in N(v)$. Particularly, messages are the result of combining the hidden states of connected nodes ($h_v$, $h_j$) with a message function $m($·$)$, which is typically approximated by a neural network and is uniformly applied over all node pairs in the graph. Then, all the messages received in a node are combined with an aggregation function $a($·$)$, producing a fixed-size output independently of the number of messages received (i.e., the number of nodes connected). This aggregation function is often implemented as an element-wise summation. 

Lastly, each node updates its hidden state ($h_v$) based on the aggregated messages received from its neighbors, using an update function $u($·$)$ also approximated by a neural network.

Formally, the message passing at a given iteration $t$ is defined as follows:
\begin{equation}
    \small m_{v,j} = m(h_v^t,\;h_j^t,\;e_{v,j})
\end{equation} 
\vspace{-0.45cm}
\begin{equation}
    \small M_v^{t+1} = a(\{m_{v,j} \;|\; j \in N(v)\})
\end{equation}
\begin{equation}
    \small h_v^{t+1} = u(h_v^t,\;M_v^{t+1})
\end{equation}

Given the final hidden states obtained after $T$ message-passing iterations, the GNN executes a readout phase. In this context, a subset of hidden-states -- which depends on the specific GNN model -- is passed through a learnable readout function $r($·$)$ that produces the output of the GNN model. Thus, $r($·$)$ is mainly intended to map the final nodes' hidden-state embeddings ($h_v^T$) to the output labels of the model $\hat{y}$:
\begin{equation}
    \small \hat{y} = r({h_v^T \;|\; v \in V})
\end{equation}

As a result, the novel message-passing architecture of GNNs endows these models with an unprecedented generalization power over graphs of different size and structure. We refer interested readers to \cite{wu2020comprehensive} for further details regarding GNNs.

\section{Proposed GNN-based NIDS} \label{sec:NIDS_architecture}

This section describes the proposed GNN-based NIDS. We first present a host-connection graph we use to represent the traffic, which has enough expressiveness to represent flow patterns of attacks, such as those depicted in Figure~\ref{fig:graph_patterns}. Then, we describe a novel GNN architecture tailored to operate over the previous graph. This new GNN model comprises a non-standard message-passing architecture that deals with the heterogeneous elements and the particularities of the network intrusion detection problem.

\subsection{Host-Connection Graph Representation}\label{sec:IP-connection_graph}
Given a set of flows $\mathcal{F}$, we build a host-connection graph $G_\mathcal{F}$, that includes a node for each distinct host involved \mbox{-- either} sending or receiving traffic. Moreover, each flow is represented as a node of this graph. Thus, given a flow $f\in \mathcal{F}$, with a source host $S$, and a destination host $D$, we create two undirected edges: one from the source host to the flow ($S \rightarrow f$), and another from the flow node to the destination host ($f \rightarrow D$). To this end, we consider IPs as identifiers for hosts. 

We build this representation based on a previous analysis on the expressiveness needed to properly capture flow patterns of attacks. Thus, this host-connection graph comprises relevant aspects of flows, with focus on their structural features. First, it enables to differentiate and relate features for the upstream and downstream traffic of the flow, and second -- and more important -- the graph explicitly represents the relations between different flows, which are connected to the same source/destination hosts. 

Note that a more straightforward representation would be to consider only hosts as node graphs, and flows as graph edges connecting the src/dst hosts. However, the decision to add specific nodes representing each flow was driven by the way GNN models operate. Note that GNNs consider only as learnable objects the hidden states of nodes in input graphs (as described in Sec.~\ref{sec:background}). As a result, to properly learn embeddings on flows, it is needed to add them as nodes of the graph. Note that this graph representation includes heterogeneous elements (i.e., hosts and flows), which is not well supported by standard GNN models. This call us to devise a new message-passing architecture specifically adapted to process and learn the host-connection graph described in this section.
\vspace{1cm}
\subsection{GNN model description}

This section describes the proposed GNN model, which comprises a non-standard message-passing algorithm that adapts to the needs of the network intrusion detection problem, considering as input the host-connection graph representation described in Section~\ref{sec:IP-connection_graph}.

\begin{figure}
    \centering
    \includegraphics[width=\linewidth]{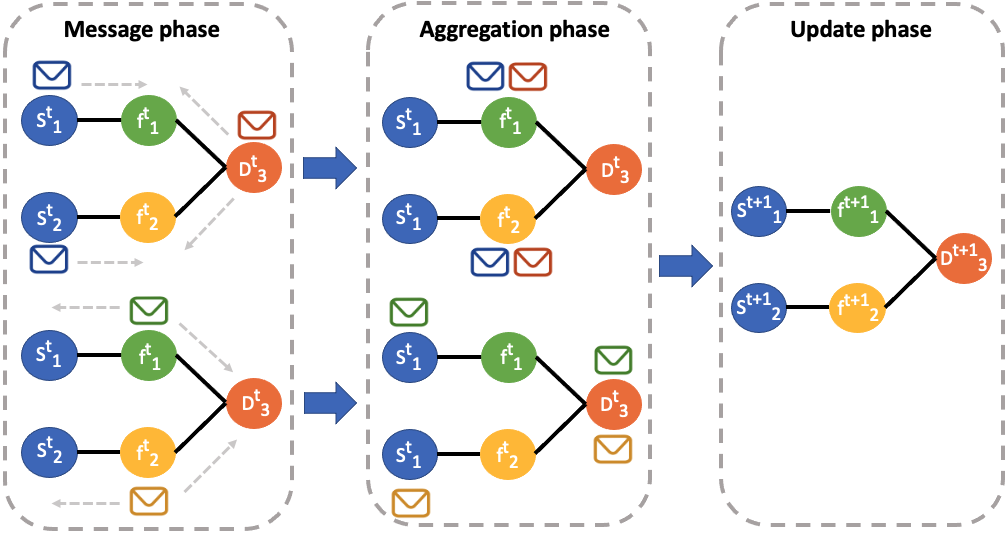}
    \caption{Illustration of the message-passing phase of the proposed GNN-based NIDS.}
    \label{fig:schema_mp_proposal}
\end{figure}

Let us define $h_i^t$ as the hidden state of node $i$ during iteration $t$, and $X_i$ as the initial features for nodes' hidden states. In the host-connection graph, nodes can represent a host, or a flow, so the hidden states of these nodes will be typically initialized with features of different nature. These features will depend on the monitoring data accessible in the network. Without loss of generality, let us assume that initial features are $ X_i = [x_0,...,x_k]$. In this case, we form the initial hidden state of flow $i$ as follows:
\begin{equation}
    \small h_i^0 = [x_0,...,x_k,0,0,0...,0]
\end{equation}

Note that hidden state vectors have a pre-defined length typically larger than the number of elements in the initial feature vector. Thus, hidden states are zero-padded.

Alternatively, if node $i$ represents a host, we simply encode all ones in the initial hidden-state $h_i^0$. In this context, it is important to avoid identifiers of nodes. A typical example would be to avoid using IP addresses to initialize hosts. This would not be desirable for the model, as the objective is to focus on the structural flow patterns, thus achieving a more general and robust characterization of attacks.

We first describe the message-passing phase of the GNN, which naturally considers the heterogeneity of the graph. Figure~\ref{fig:schema_mp_proposal} illustrates the message-passing process. Formally, we apply the following operation in each message-passing iteration $t \in [T]$:

\begin{equation}
    \label{eq:message_phase}
    \small a_i^t = \frac{1}{|\mathcal{N}(i)|} \sum_{j \in \mathcal{N}(i)} \sigma_{type}(h_i^t \;||\; h_j^t)
\end{equation}
\begin{equation}
    \label{eq:update_phase}
    \small h_i^{t+1} = \delta_{type}(h_i^t \;||\; a_i^t)
\end{equation}

First, the model applies a learnable message function $\sigma_{type}$ given the concatenation of the hidden states of two connected nodes --i.e., an edge in the input graph of the GNN. Here, $\sigma_{type}$ implicitly comprises two possible functions, which respectively depend on the type of edge where they are applied: $\sigma_{sf}$ for edges ($S \rightarrow f$), and $\sigma_{fd}$ for edges ($f \rightarrow D$), according to the description of the host-connection graph in Section~\ref{sec:IP-connection_graph}.

Afterward, an aggregation function is applied to the messages computed on each node. For this, we apply an element-wise mean over messages. In our case, using this function helps better normalize data across the multiple message-passing iterations, rather than using a standard element-wise summation.

Finally, the hidden states are updated considering the information collected in the new aggregated message. This is done by applying the update function $\delta_{type}$ to the aggregated message and the current hidden state of the node. Similarly to the message function, $\delta_{type}$ comprises two different learnable functions ($\delta_{h}$ and $\delta_{f}$) respectively applied to update the hosts' and the flows' hidden states.

As a result, the $\sigma_{sf}$, $\sigma_{fd}$, $\delta_{h}$ and $\delta_{f}$ functions are all learnable functions than can be approximated by neural networks during training. Particularly, we implement $\sigma_{sf}$ and $\sigma_{fd}$ as 2-layer fully-connected NNs, while $\delta_{h}$ and $\delta_{f}$ are modeled as Gated Recurrent Units (GRUs~\cite{chung2014empirical}).

Finally we define the readout function $r($·$)$ as follows: 
\begin{equation}
    \small y_i = r(h_{i}^{T})
\end{equation}

The function $r($·$)$ takes as input the final hidden states of each flow, and outputs the predicted class for the flow (either a specific attack, or benign traffic). This function is implemented with a 3-layer fully-connected NN, where all the possible output classes are represented via one-hot encoding.

We use \textit{ReLU} activation functions on all the layers of the different NNs mentioned above. Except for the last layer of the $r($·$)$ function, which uses a \textit{softmax} activation. As we apply this GNN model for multi-class classification, we use a \textit{categorical cross-entropy} loss function for training. However, the model could also be directly used for binary classification (e.g., classify flows on malicious or benign traffic) using a \textit{binary cross-entropy} loss function instead.

\section{Evaluation}

This section presents an evaluation of the proposed GNN-based NIDS, following two main directions. First, we evaluate the accuracy of the system compared to other state-of-the-art ML-based NIDS, using the well-known CIC-IDS2017 dataset~\cite{sharafaldin2018toward}. Then, we artificially generate some common adversarial attacks in the previous dataset, to analyze the robustness of our GNN model compared to the other ML-based benchmarks.

\subsection{Dataset}\label{sec:datasets}

To evaluate the proposed GNN-based NIDS -- described in Sec.~\ref{sec:NIDS_architecture}--, we use the well-known CIC-IDS2017~\cite{sharafaldin2018toward}, which contains a representative collection of up-to-date attacks well mixed with real-world traffic. More in detail, malicious traffic is classified in 7 broad classes of attacks: Brute Force, Heartbleed, Botnets, Dos, DDos, Infiltrations and Web attacks. In total, there are 15 different sub-classes of attacks. Likewise, each flow record aggregates a total of 80 features. We refer interested readers to~\cite{sharafaldin2018toward} for further details on this dataset.

Moreover, we evaluate our model with a training and validation dataset generated through a random split of 80\% and 20\% graph samples respectively -- totaling 895,400 flows for training, and 223,850 flows for validation. In our experiments, we show the results averaging over 5 cross-validations following the aforementioned evaluation methodology. 

\subsection{Experimental results of the NIDS} \label{sec:expers}

This section evaluates the accuracy of the proposed GNN-based NIDS over the CIC-IDS2017 dataset (Sec. \ref{sec:datasets}).

A main difficulty when training ML-based NIDS is that datasets are inherently imbalanced, having a great bulk of benign traffic and a small portion of traffic related to attacks. For instance, in the CIC-IDS2017 dataset, malicious traffic represents only $\approx$12\% of the flows. To address this, we first make make a pre-processing of the dataset. We randomly drop $90\%$ of the graphs containing only normal traffic (Benign class), thus over-representing traffic belonging to attacks. For the evaluation, we consider only classes with more than 100 flow samples, resulting in 12 different sub-classes. We tested the use of loss functions specifically designed for imbalanced datasets (e.g., Focal loss~\cite{lin2017focal}), however we did not find a significant improvement, finally using a common categorical cross-entropy loss function. 

Table~\ref{tab:results_ids2017} summarizes the accuracy achieved by our GNN-based NIDS with respect to a collection of ML methods commonly used in state-of-the-art NIDS. In particular, we benchmark our solution against a 3-layer Multilayer perceptron (MLP)~\cite{gardner1998artificial}, Ada-boost~\cite{hastie2009multi}, Random Forest (RF)~\cite{biau2016random} and ID3~\cite{hssina2014comparative}. We use a standard weighted F1-score to measure the per-class accuracy, which unifies in a single metric the precision and recall of solutions. From these results, we can observe that the proposed model achieves a level of accuracy comparable to state-of-the-art ML methods, obtaining a weighted F1-score of 0.99 over all traffic flows.

\begin{table}
\centering
\caption {Accuracy of different ML-based NIDS over the CIC-IDS2017 dataset (11 attack classes + Benign traffic).}
\resizebox{\columnwidth}{!}{%
\begin{tabular}{c| c c c c c}
\textbf{Class label} & \textbf{MLP} & \textbf{AdaBoost}&  \textbf{RF} & \textbf{ID3}& \textbf{Our proposal}\\
\hline
\hline
Benign & 0.67 &0.68 & \textbf{0.99}  & \textbf{0.99} & \textbf{0.99} \\ 
\hline
SSH-Patator & 0.0 &0.0 & \textbf{0.99} & \textbf{0.99} & 0.98\\  
 \hline
FTP-Patator& 0.0 &0.0 &\textbf{0.99} &\textbf{0.99} & \textbf{0.99} \\ 
 \hline
DoS GoldenEye & 0.12 &0.0 &0.97 &0.96 & \textbf{0.99} \\ 
 \hline
DosHulk & 0.63 &0.63 & \textbf{0.99}&\textbf{0.99} & \textbf{0.99}\\ 
 \hline
DoS slowloris& 0.02 &0.0 & \textbf{0.99}& \textbf{0.99} & 0.98\\
\hline
DoS Slowhttptest& 0.01 &0.0 & \textbf{0.98} & \textbf{0.98} & 0.97 \\  
 \hline
DDoS& 0.51 &0.0 & \textbf{0.99} & \textbf{0.99} & \textbf{0.99} \\ 
 \hline
Web-Brute Force & 0.0 & 0.0 & \textbf{0.82} & 0.76& 0.73\\ 
 \hline
Web-XSS &0.0 &0.0 &0.69  &0.65 & \textbf{0.83}\\
 \hline
Bot & 0.0 & 0.0 & \textbf{0.98}  & \textbf{0.98} & \textbf{0.98}\\
 \hline
Port Scan & 0.78 &0.0 & \textbf{0.99}  & \textbf{0.99} & \textbf{0.99} \\
 \hline
\end{tabular}
}
\vspace{-0.2cm}
\label{tab:results_ids2017}
\end{table}

\subsection{Robustness Against Adversarial Attacks}

\begin{figure}[!t]
    \centering
    \includegraphics[width=0.9\linewidth]{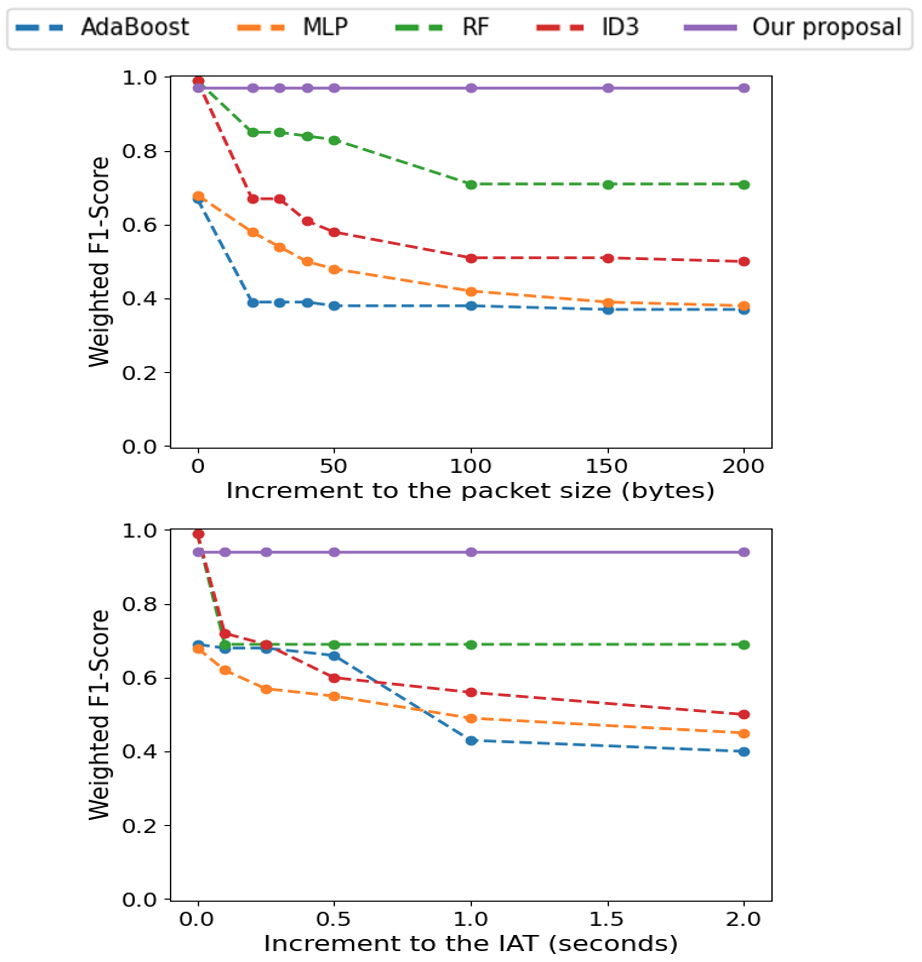}
    \vspace{-0.2cm}
    \caption{Evaluation of the accuracy of ML-based NIDS under variations of the packet size (top) and the inter-arrival time (bottom).}
    \vspace{-0.3cm}
    \label{fig:generalization_power}
\end{figure}

Section~\ref{sec:expers} shows that fullthe proposed GNN-based NIDS achieves similar accuracy to state-of-the-art ML-based NIDS over the CIC-IDS2017 dataset. Nevertheless, in Section~\ref{sec:graph_learning} we discussed the limitations of traditional flow-based ML-based methods, which can be highly vulnerable to variations in individual flow features. This section aims to analyze the robustness of the proposed NIDS when facing common adversarial attacks~\cite{corona2013adversarial}.

Typically, the packet size is a highly discriminative feature for detecting many attacks at the level of individual flows (e.g. DDoS). Consequently, in our first experiment, we artificially increment the packet size of attack-related flows, in order to test the robustness of the proposed NIDS against this potential detection prevention method. Figure~\ref{fig:generalization_power} (top) shows the evolution of the accuracy (F1-score) as the packet size of attack-related flows is incremented ([0, 200] bytes). As we can observe, the proposed GNN model is able to keep the same level of accuracy as in the experiments of the previous section, showing robustness to this adversarial attack. In contrast, the other ML benchmarks suffer a significant degradation on their accuracy as the packet size increases.

In a second experiment, we propose to make variations on the throughput of attack-related flows. For this, we artificially increase the inter-packet arrival times, to serve traffic at lower rates. Thus, Figure~\ref{fig:generalization_power} (bottom) shows the results after applying increments to inter-arrival times ([0, 2] seconds), showing again that the proposed model maintains its base level of accuracy; while the other ML benchmarks considerably decrease their accuracy (up to 50\%).

Overall, we can observe in Figure~\ref{fig:generalization_power} that all the baseline ML models evaluated exhibit strong vulnerability against common adversarial attacks that modify flow-level features. In contrast, the proposed GNN model maintains the same level of accuracy, being completely robust to these detection prevention methods. This is mainly due to its ability to capture the structural flow patterns of attacks, which remain unchanged even after varying flow features.

\section{Related Work}

The application of ML for intrusion and anomaly detection has been largely investigated by the research community. First ML-based solutions proposed the use of traditional ML algorithms, such as K-nearest neighbors (KNN), Support Vector Machines (SVM), Random Forests (RF), or a combination of them, to classify network attacks.
There are numerous surveys that cover the vast related work in this area (e.g.,~\cite{khraisat2019survey, resende2018survey,ghorbani2009network, garcia2009anomaly}).

More recent works propose also the use of Deep Learning techniques (i.e., neural networks). For instance,~\cite{he2019novel} proposes a NIDS that combines deep Autoencoders and Long Short-Term Memory (LSTM) cells. In this system, the Autoencoder learns relevant flow embeddings, which are then fed to a LSTM model to classify the attacks on flows, by opportunistically exploiting the temporal dependencies in the data. Other works such as~\cite{hassan2020hybrid} propose the use of Convolutional Neural Networks (CNN) to extract meaningful information from NIDS data.

Many ML-based NIDS treat and classify traffic flows independently, which is a limiting factor for the detection of common multi-flow attacks (e.g., DDoS, network scans)~\cite{khraisat2019survey}. As a result, few recent works have explored the aggregation of network traffic into clusters (e.g.,~\cite{garcia2014empirical}) or graphs. In particular,~\cite{xu2021detecting, busch2021nf} propose the use of graph learning to exploit the relationship among network connections, showing significant improvement for malware detection in mobile applications. Similarly,~\cite{zhou2020automating} approaches the problem of Botnet detection assuming visibility of the full botnet topology. However, none of these previous works show robustness against adversarial attacks that produce variations on flow features to evade detection, such as the packet size, or throughput (inter-packet arrival times).

Recently, some pioneering works started to unveil the potential of GNNs for other networking problems, such as network modeling~\cite{rusek2019unveiling, rusek2020routenet}, network optimization~\cite{almasan2019deep}, network planning~\cite{hang2021network} or network troubleshooting~\cite{meng2020interpreting, pujol-perich2021netxplain}. In this work, we show that GNNs can also represent a breakthrough in the field of network intrusion detection.

\section{Conclusions} 
In this paper we motivate the use of Graph Neural Networks (GNNs) to develop accurate and robust NIDS. We argue that, to achieve effective ML-based NIDS, it is essential not only to collect relevant patterns on individual flow features, but also to capture meaningful structural flow patterns that characterize different attacks (e.g., DDoS, port/network scans, brute force attacks). To this end, we first present a graph representation that properly represents the properties of flows and their relationships in the network. Then, we present a novel GNN architecture specifically designed to learn and generalize over the previous graph-structured information. 

First, we have tested the accuracy of our model, showing comparable results to state-of-the-art ML-based NIDS in the well-known CIC-IDS2017 dataset. Then, we have tested the solution against two common adversarial attacks that intentionally modify relevant flow features on attack-related flows (packet size and inter-arrival times) to evade detection. The results show that, while the proposed GNN model is completely robust to these attacks, state-of-the-art ML models for NIDS degrade their accuracy up to 50\%. This is mainly thanks to the capability of the proposed GNN model to \textit{learn} the inherent structural flow patterns that compose different attacks. These structural patterns represent a deeper knowledge about attacks, as they typically remain unchanged over time, and across different networks.

\section*{Acknowledgements}
This work has received funding from the European Union’s Horizon 2020 research and innovation programme within the framework of the NGI-POINTER Project funded under grant agreement No. 871528. This paper reflects only the author's view; the European Commission is not responsible for any use that may be made of the information it contains. This work was also supported by the Spanish MINECO under contract TEC2017-90034-C2-1-R (ALLIANCE) and the Catalan Institution for Research and Advanced Studies (ICREA).

\balance
\def\newblock{\ }%
\bibliographystyle{ACM-Reference-Format}
\bibliography{references}

\end{document}